\documentclass[12pt]{article}

\usepackage{epsfig}

\newcommand{\labl}[1]{\label{#1}}
\newcommand{\bo}[1]{\bar{\Omega}_{#1}}
\newcommand{\oo}[1]{\Omega_{#1}}
\newcommand{\eq}{\begin{equation}}
\newcommand{\eqx}{\end{equation}}
\newcommand{\eqn}{\begin{eqnarray}}
\newcommand{\eqnx}{\end{eqnarray}}
\newcommand{\f}[2]{\frac{#1}{#2}}

\begin{document}

\begin{center}
{\Large\bf  SOLUTION OF THE ODDERON PROBLEM  }\\
\vspace*{2cm}
{\large\bf R. A. Janik and J. Wosiek}  \\
\mbox{ }\\
Institute of Physics, Jagellonian University \\
Reymonta 4. 30-059 Cracow, Poland \\
\end{center}
\vspace*{2cm}

\begin{abstract}
The intercept of the odderon trajectory is derived, by finding 
the spectrum of the second integral of motion of the three reggeon
system in high energy QCD. When combined with earlier solution of the
appropriate Baxter equation, this leads to the determination
of the low lying states of that system. In particular,
the energy of the lowest state gives
 the intercept of the odderon 
$\alpha_O(0)=1-0.2472\alpha_s N_c/\pi$.

\end{abstract}
 
\vspace*{2cm}
\noindent TPJU-2/98 \newline
February 1998 \newline
hep-th/9802100
\newpage
One of the, still unsolved, problems of perturbative QCD is the behaviour
of the 
theory in the Regge limit, characterized by  high energies and fixed 
momentum transfers.
This limit --- the (Generalized) Leading Logarithmic
Approximation --- is described by the exchange of
reggeized gluons. 
 The intercept of the pomeron trajectory, the BFKL
pomeron, has been derived in the classic works of Balitski, Fadin,
 Kuryaev and Lipatov \cite{BFKL}. The
next natural step is to find the intercept of the odderon
trajectory, which however, turned out to be very difficult
\cite{firstb,LipOd,GLN}. 

An important progress was made by Lipatov, Faddeev and Korchemsky 
\cite{LIP0,FK} who reduced the problem to the solution of a functional 
equation --- the Baxter equation --- for physical values of the two relevant 
constants of
motion ($q_2$ and $q_3$) of the system of three reggeized gluons. 
Various approximation techniques \cite{KorQ,Wallon,KorWh}
for solving the Baxter equation have been used, and in our previous
work \cite{USPRL,USPROC} an exact method of  constructing a solution
for general $q_2$ and $q_3$ was developed. However, while the eigenvalues 
of $\hat{q}_2$ 
are known, the  spectrum of $\hat{q}_3$
remained unavailable, apart from asymptotic results of
\cite{KorQ,KorWh}. In this letter we report on the solution
of the eigenproblem of $\hat{q}_3$
which removes the last obstacle in deriving the numerical value of $\alpha_O$
in the Leading Logarithmic Approximation. 

The intercept of the odderon trajectory is given by 
\begin{equation}
\alpha_O(0)=1+{\alpha_s N_c \over 4\pi}\left(\epsilon_3(h,q_3)+
 \overline{\epsilon}_3(\overline{h},\overline{q}_3)\right), \labl{inter}
\end{equation}
where $\epsilon_3$ and $\overline{\epsilon}_3 $ are respectively the largest 
eigenvalues
of the three reggeon hamiltonian and its antiholomorphic counterpart
 \cite{FK}. The conformal weight, $h$, parameterizes the eigenvalues
of the Casimir operator $\hat{q}_2$
\begin{equation}
q_2=h(1-h),\;\;\;\;h={1\over 2}(1+m) -i\nu,\;\;  m\in Z, \nu\in R.
\labl{q2}
\end{equation}
Analogous formulas hold for the antiholomorphic sector
with $\overline{h}=(1-m)/2-i\nu$ \cite{FK}. 
After an explicit expression for the energies
$\epsilon_3(h,q_3)$ ($\overline{\epsilon}_3(\bar{h},\bar{q}_3)$) 
 $q_2$ and $q_3$ was derived \cite{USPRL} the only unknown ingredient was 
the quantization of $q_3$.

The eigenproblem of the $\hat{q}_3$ operator was formulated 
in general terms by Lipatov \cite{LipOd}. However the quantitative 
solution was lacking due to the complicated and indirect way 
in which the 
boundary conditions enter and fix the spectrum. To begin, we quote 
a general form of the wave function of the compound state of three
reggeized gluons
\begin{equation} 
f(\rho,\bar{\rho})=\sum_{r,s} c_{rs} f_r(\rho) \bar{f}_s(\bar{\rho}),
\labl{odfun} 
\end{equation}
where $\rho=\{\rho_1,\rho_2,\rho_3\}$ denotes three transverse
coordinates in the complex number representation $\rho_k=x_k+i y_k,\;
k=1,2,3$. All other quantum numbers are collectively denoted by index
$r$. Similarly in the antiholomorphic sector $\bar{\rho}_k=x_k-i y_k$.
Since the integrals of motion $\hat{q}_2, \hat{q}_3$ commute with the
hamiltonian, one chooses $f(\rho)$ as a simultaneous eigenfunction of
these operators.

With the conformally covariant Ansatz, 
$z={\rho_{12}\rho_{30}\over\rho_{10}\rho_{32}}$, $\mu=h/3$,
\begin{equation}
f_{\rho_0,q_2,q_3}=\left({\rho_{12}\rho_{13}\rho_{23}\over 
\rho_{10}^2 \rho_{20}^2 \rho_{30}^2}
\right)^{\mu} \Phi^{(h,q_3)}(z),   \labl{ansatz}
\end{equation}
the eigenequation, $\hat{q}_3 f = q_3 f$, reads, in terms of $\Phi$, 
and at fixed $q_2$ given by (\ref{q2}).
 \begin{equation}
a(z) {d^3 \over d z^3}\Phi(z) + b(z){d^2\over d z^2} \Phi(z)
+c(z){d\over dz}\Phi(z) + d(z) \Phi(z)=0,  
  \labl{diff} 
\end{equation}
where
\begin{eqnarray}
a(z)=z^3(1-z)^3=\sum_{i=0}^3 a_i z^{i+3},\; 
b(z)=2z^2(1-z)^2(1-2z)=\sum_{i=0}^3 b_i z^{i+2},&&\nonumber\\
c(z)= z(z-1)\left( z(z-1)(3\mu+2)(\mu-1)+3\mu^2-\mu\right)
=\sum_{i=0}^3 c_i z^{i+1},&&\nonumber \\
d(z)= \mu^2(1-\mu)(z+1)(z-2)(2z-1) -i q_3 z (1-z)=\sum_{i=0}^3 d_i z^{i}.&&
\nonumber\end{eqnarray}

This is a third order linear differential equation 
with the three regular singular points at $z=0,1$ and $\infty$,
introduced by Lipatov in \cite{LipOd}, and investigated in \cite{jan2}
in a slightly different form. We will solve this 
equation by standard methods and identify proper boundary conditions
which lead to the quantization of $q_3$. To this end
we first construct a fundamental set of three linearly independent
solutions 
$
\vec{u}^{(0)}(z) = (u_1^{(0)}(z),u_2^{(0)}(z),u_3^{(0)}(z)),
$
around $z=0$.   
\begin{eqnarray}
u_1^{(0)}(z)=z^{s_1} \sum_{n=0}^{\infty} f_n^{(1)} z^n,\;
u_2^{(0)}(z)=z^{s_2} \sum_{n=0}^{\infty} f_n^{(2)} z^n,&& \labl{sep} \\
u_3^{(0)}(z)= \log{z}\; u_2^{(0)}(z)+ 
          z^{s_3} \sum_{n=0}^{\infty} r_n z^n,
         \nonumber   
\end{eqnarray} 
where $s_1=2h/3,\; s_2=-h/3+1,\; s_3=-h/3$ and the
 coefficients of the expansions are determined by the following
recursion relations 
\begin{eqnarray}
f_{n}^{(i)}=-\sum_{m=1}^{3} f_{n-m}^{(i)}\lambda_m(s_i+n)/\lambda_0(s_i+n),
 \;\; f_0^{(i)}=1,\;f_{-j}^{(i)}=0,\; i=1,2,j>0,\labl{rec0} \\
\lambda_m(x)=a_{m}(x-m)(x-m-1)(x-m-2)
+b_{m}(x-m)(x-m-1)&&\nonumber\\
+c_{m}(x-m)+d_{m} ,&&\nonumber
\end{eqnarray}
and for the logarithmic solution
\begin{eqnarray}  
r_{n}=-(p_{n-1}+\sum_{m=1}^3 r_{n-m}\lambda_m(s_3+n))/\lambda_0(s_3+n),&&
\nonumber \\
r_1=1,\;r_0=-p_0/\lambda_1(s_3+1),\;r_{-1}=0,\;
p_n=\sum_{m=0}^3 f^{(2)}_{n-m}\gamma_m(s_2+n), &&\labl{recl}\\
\gamma_m(x)=a_{m}\left(3(x-m)(x-m+2)+2\right)+b_{m}\left(2(x-m)-1\right)
+c_{m}.&&\nonumber
\end{eqnarray}
There are two physical conditions which our solution should satisfy
 a) the complete {\em wave function} of the compound system (\ref{odfun}) 
must be single valued in the whole transverse plane of reggeon coordinates,
and
 b) the wave function (\ref{odfun}) must obey Bose symmetry.
 Together with the analyticity of the solution $\Phi(z)$ these conditions
 unambiguously determine the spectrum of $\hat{q}_3$.

The series in Eq.(\ref{sep}) are convergent in the unit circle $R_0$ around
$z=0$, and therefore they determine uniquely the analytic continuation
$\vec{u}^{(0)}(z)$ to the cut complex plane. 
To achieve this continuation in practice we
construct two other fundamental sets of solutions
$\vec{u}^{(1)}(z)$ and $\vec{u}^{(\infty)}(z)$ around $z=1$ and $z=\infty$.
 This could be done analogously to Eqs.(\ref{sep}-\ref{recl}),
for the transformed equation,
 however, because of the symmetry under permutations of reggeon coordinates, 
one can
partly satisfy condition b) by a proper choice of these bases. We define
\begin{equation}
\vec{u}^{(1)}(z)=\vec{u}^{(0)}(1-{1\over z}),\quad\quad\quad\quad
\vec{u}^{(\infty)}(z)=\vec{u}^{(0)}({1\over 1-z}).
\label{e.basis}
\end{equation}
These series are convergent in the regions $R_1: Re(z) > 1/2$
and \mbox{$R_{\infty}: |1-z|>1$} respectively. 
Analytic continuation is realized 
by the transition matrices
\begin{equation}
{u}^{(0)}_i(z)=\Gamma_{ij}{u}^{(1)}_j(z).\quad\quad\quad\quad
{u}^{(1)}_i(z)=\Omega_{ij}{u}^{(\infty)}_j(z), \labl{tran}
\end{equation}
which depend on $h$ and $q_3$ only.
They contain the full information
about the system, in particular about its spectrum.
In practice we calculate transition matrices as the solutions of the systems 
of $3\times 3$
algebraic equations (\ref{tran}) written at some judiciously chosen point 
$z=\zeta$. For example
\begin{equation}
\Gamma_{ik}={W_{ik}\over W(\vec{u}^{(1)}(\zeta))},\;\;
W_{ik}=W(u^{(1)}_k\rightarrow u^{(0)}_i), \labl{wron}
\end{equation}
where $W(\vec{u}^{(1)}(\zeta))$ is the Wro\'{n}ski determinant of the 
fundamental solutions $\vec{u}^{(1)}(\zeta)$. As long as $\zeta$ is in the
intersection of the convergence regions $R_0$ and $R_1$, the matrix elements
obtained from Eq.(\ref{wron}) are independent of $\zeta$ provided
enough terms in the series (\ref{sep}) are included. 

Finally we implement the uniqueness constraints a). It is crucial
to observe that requiring singlevaluedness in the holomorphic and 
antiholomorphic sectors separately gives a too strong condition
and is in fact not necessary. Even though the hamiltonians in both
sectors commute, 
proper boundary conditions should be formulated only for the
 wave function of the whole system (\ref{odfun}). We therefore
define a general bilinear form \footnote{The power prefactors 
in (\ref{ansatz}) are irrelevant for this discussion.}
\begin{equation}
\label{psisum}
\Psi_{h,\bar{h},q_3,\bar{q}_3}(z,\bar{z})
={\vec{\bar{u}}^{(1)}(\bar{z})}^T A^{(1)} \vec{u}^{(1)}(z), \labl{com}
\end{equation} 
and demand its uniqueness in the whole transverse plane. The
compound function 
(\ref{com}) has 9 free parameters \footnote{We thank Gregory Korchemsky
for the discussion on that point.}. 
By inspecting Eqs.(\ref{sep}) we see that the most general
choice of coefficients, consistent with the uniqueness of the wave
function in the neighbourhood of $z=1$, is
\begin{equation}
 A^{(1)}=\left(    \begin{array}{ccc}
              \alpha & 0     & 0           \\
              0      & \beta & \gamma      \\  
              0      & \gamma& 0           \\
              \end{array} \right),  \labl{uniq}
\end{equation}
which has freedom of three parameters.
Incidentally this form guarantees also the normalizability of the wave function.
 Rewriting the wave function in
terms of other bases (\ref{e.basis}) around $z=0$ and $z=\infty$, gives 
another coefficient matrices  
\begin{equation}
A^{(0)}=(\bar{\Gamma}^{-1})^T A^{(1)} \Gamma^{-1},\; 
A^{(\infty)}=\bar{\Omega}^T A^{(1)} \Omega. \labl{fix} 
\end{equation}
Now, uniqueness in the whole
transverse plane requires that the transformed matrices, Eq.(\ref{fix})
have the same form as (\ref{uniq}), with possibly different coefficients.
Therefore we require
\begin{eqnarray}
A^{(0)}_{12}=A^{(0)}_{13}&=A^{(0)}_{33}=&A^{(0)}_{21}=A^{(0)}_{31}=0,
                        \labl{a0}  \\
A^{(\infty)}_{12}=A^{(\infty)}_{13}&=A^{(\infty)}_{33}=&
A^{(\infty)}_{21}=A^{(\infty)}_{31}=0. \labl{ai}
\end{eqnarray}
In fact only one of these sets is sufficient, as can be seen from the following
topological argument. Any possible cut in the domain of the full wave 
function has
to begin and end at the singular points of the equation, i.e. at $0,1$
or at $\infty$. Therefore eliminating two of these points guarantees
that there is no cut beginning at the third one. Equations
(\ref{a0},\ref{ai}) are linear homogeneous equations for the
coefficients $\alpha,\beta$ and $\gamma$.  
The condition of the existence of a nonzero solutions of Eqs(\ref{a0})
or (\ref{ai}) provides the quantization of $q_3$ and $\bar{q}_3$ we
looked for. 
It can conveniently be written as
\eq
{\cal B}_U \left(\begin{array}{c} \alpha\\ \beta\\\gamma\\
\end{array}\right) =0 \quad\quad and \quad\quad
{\cal B}_L \left(\begin{array}{c} \alpha\\ \beta\\ \gamma\\
\end{array}\right) =0 .
\label{qond}
\eqx
where the rows of the matrix ${\cal B}_U$ (${\cal B}_L$) are the
coefficients of $\alpha$, $\beta$, $\gamma$ in, e.g., $A^{(\infty)}_{12}$,
$A^{(\infty)}_{13}$ and $A^{(\infty)}_{33}$ ($A^{(\infty)}_{21}$,
$A^{(\infty)}_{31}$ and $A^{(\infty)}_{33}$). Explicitly 
\eq
 {\cal B}_U=\left(    \begin{array}{ccc}
  \bo{11}\oo{12} & \bo{21}\oo{22} & \bo{21}\oo{32}+\bo{31}\oo{22}\\
  \bo{11}\oo{13} & \bo{21}\oo{23} & \bo{21}\oo{33}+\bo{31}\oo{23}\\
  \bo{13}\oo{13} & \bo{23}\oo{23} & \bo{23}\oo{33}+\bo{33}\oo{23}\\

              \end{array} \right).
\labl{bs} 
\eqx

    Figure 1 shows the, suitably transformed, absolute value of the determinant of
${\cal B}_U$ as a function of $q_3$ along the imaginary axis in the complex $q_3$ plane.
 The eigenvalue
 of the first Casimir operator $q_2$ is fixed to $q_2=\bar{q}_2=1/4$
which corresponds to the lowest representation of the $SL(2,C)$.
All formulae in the antiholomorphic sector are the same with 
$\bar{q}_3=q_3^{\star}$, a $\star$ denoting a complex conjugate \cite{FK}.
Clearly a set of discrete $q_3$ values exists where the first condition
 (\ref{qond}) is satisfied. Imposing the second condition eliminates 
half of the candidates. In addition a discrete series of real
solutions of (\ref{qond}) exists.
 Both groups lead to single valued wave
functions, however the condition of Bose 
symmetry singles out the imaginary $q_3$ only. Namely, for purely
imaginary eigenvalues, the matrices $A^{(0)}$, $A^{(1)}$ and $A^{(\infty)}$
coincide. This, together with the definition of the basis
(\ref{e.basis}), guarantees the invariance of the wave function under
even permutations. In order to implement a full Bose symmetry it
suffices to take the complete wave function as
\eq
\Psi(z,\bar{z})=\Psi_{\f{1}{2},\f{1}{2},q_3,\bar{q}_3}(z,\bar{z})+
\Psi_{\f{1}{2},\f{1}{2},q_3,\bar{q}_3}(\f{z}{z-1},\f{\bar{z}}{\bar{z}-1})
\eqx
The second term is just the wave function in the $(-q_3,-\bar{q}_3)$
sector which, due to the degeneracy $q_3\longleftrightarrow  -q_3$, 
 obeys the general structure (\ref{odfun}).
On the other hand, for real eigenvalues $q_3$,  the matrices 
$A^{(0)}$, $A^{(1)}$
and $A^{(\infty)}$ differ by a phase factor $e^{2\pi i/3}$, and the only
symmetric solution is identical with zero.  

For $q_3$ lying outside the real and imaginary axes 
{\em both} constraints (\ref{qond}) cannot be satisfied simultaneously. 
Therefore we are led to conclude that the physical spectrum of 
$q_3$ for $h=1/2$ lies on the imaginary axis. 
The first few levels are quoted in Table 1.
 
   It is very instructive to superimpose this result on our 
earlier calculations, based on a different approach (Bethe Ansatz), 
which resulted
in the analytic expression for the eigenenergy of the three reggeon
system as a function of $h$ and $q_3$ \cite{USPRL}. Figure 2 shows 
$\epsilon_3(1/2,q_3)$ along the imaginary axis of $q_3$. Black dots
and crosses 
mark values of $q_3$ quantized according to the first condition in
(\ref{qond}). It turns out that the candidates which were eliminated 
by the second condition (crosses) are numerically very close to 
the poles of the $\epsilon_3$. They are, however, non physical 
since the corresponding wave functions are not single valued. 

   The intercept of the odderon trajectory is determined by the 
largest eigenvalue $\epsilon_3(1/2,q_3^O)$. This corresponds to
the first non-zero $q_3^O=q_3^{(1)}$ on the imaginary axis, 
with the numerical value 
\begin{equation}
q_3^O=-0.20526i,
\end{equation}
which, together with our solution of the Baxter Equation, 
$\epsilon_3(h,q_3)$ \cite{USPRL}, gives for the energy 
of the odderon state
\begin{equation}
\epsilon_3(1/2,q_3^O)=-0.49434.
\end{equation}
This translates for the intercept of the odderon trajectory, 
c.f. Eq.(\ref{inter}),
\begin{equation}
\alpha_O(0)=1-0.24717{\alpha_s N_c\over\pi},  \labl{fin}
\end{equation}
which may solve the longstanding phenomenological puzzle why
the odderon trajectory is so hard to observe experimentally. However
any phenomenological consequences of this result should be taken
with a great caution. Higher order and running $\alpha_s$
corrections can change this number. Connection between hard 
and soft exchanges should be better understood. In any case
the 
general assumptions behind the derivation of (\ref{fin}) are the same
as those of the famous calculation of the pomeron trajectory 
 in the classic work of Balitski, Fadin, 
Kuryaev and Lipatov\cite{BFKL}.

    In Table 1 we quote first few quantized values of $q_3$,
together with corresponding energies. Indeed next states have a  
substantially smaller intercept and consequently, their contribution  
to the high energy scattering is negligible. There exists also
the consistent solution of Eqs.(\ref{qond}) with $q_3=0$ which 
has smaller $\epsilon_3$ but relatively close to the energy 
of the odderon solution. However this solution is highly 
pathological and will not be discussed here in detail.   
    
    Our method provides also explicit expressions for the wave 
functions of the compound states. At the eigenvalues of $\hat{q}_3$
the coefficients of the expansion (\ref{com}) are given by the common
eigenvector corresponding to the zero eigenvalue of ${\cal B}_U$
or ${\cal B}_L$. Therefore the wave functions are 
given explicitly, in terms of known bases, and can be used for 
various applications. For the odderon state we obtain
\begin{equation}
\alpha^O=0.7096,\;\;\beta^O=-0.6894,\;\;\gamma^O=0.1457.
\end{equation}
Note that the asymptotic form of the wave function at $z=1$,
implied by the uniqueness condition (\ref{uniq}), agrees with that
derived by Lipatov from the symmetry considerations \cite{LipOd}.
Moreover, it follows from Eqs.(\ref{a0},\ref{ai}) that the 
same asymptotics holds around other singular points.
 
   A variational estimate of the lower bound for the odderon
 $\alpha_O(0) > 1+ 0.36 \alpha_s N_c/\pi$ was derived in 
Ref.\cite{GLN}. Recently this has been challenged by Braun, who 
gives the bound $\alpha_O(0)>1-.339\alpha_s N_c/\pi$ \cite{Brau}.
 The latter estimate is consistent with our exact result. 
Note that this bound is rather close to the intercept of
 our degenerate solution of the Baxter equation for $q_3=0$. 
It would be interesting to repeat their variational calculation with our
exact wave function, especially including more terms in the
 `logarithmic part' of their wave function.

Very recently Korchemsky has studied dependence of the eigenvalues
of $\hat{q}_3$ on $h$ using the sophisticated technique of Whitham dynamics
\cite{KorWh}. 
His results, as based on the quasiclassical approximation, should
agree with ours for higher states. In fact we have found earlier
that his WKB formulae reproduce exact results quite well
even at low values of $q_3$.

  \begin{table}    
  \begin{center}
   \begin{tabular}{|c|cc|cc|} \hline\hline
   {\em No.} & \multicolumn{2}{c|}{ $q_3$ } 
                                     & \multicolumn{2}{c|}{ $\epsilon_3$} \\
   \hline
  0 & \multicolumn{2}{c|}{$0$}
                                     & \multicolumn{2}{c|}{$-0.73801$} \\

  1 & \multicolumn{2}{c|}{$0.20526 i$}
                                     & \multicolumn{2}{c|}{$-0.49434$} \\
  2 & \multicolumn{2}{c|}{$2.34392 i$} 
                                    & \multicolumn{2}{c|}{$-5.16930$} \\
  3 & \multicolumn{2}{c|}{$8.32635 i$} & \multicolumn{2}{c|}{$-7.70234$}  \\ 
   \hline\hline  
   \end{tabular}
  \end{center}
\caption{Quantization of $q_3$ and corresponding eigenvalues of the holomorphic
hamiltonian.}
   \end{table}

\vspace*{1cm}
  We thank G. Korchemsky for interesting discussions.
  This work is  supported by the Polish Committee for Scientific Research 
  under grants no. PB 2P03B19609 and PB 2P03B04412.

\section*{Figures}

\begin{figure}[htb]
\epsfig{width=12cm,file=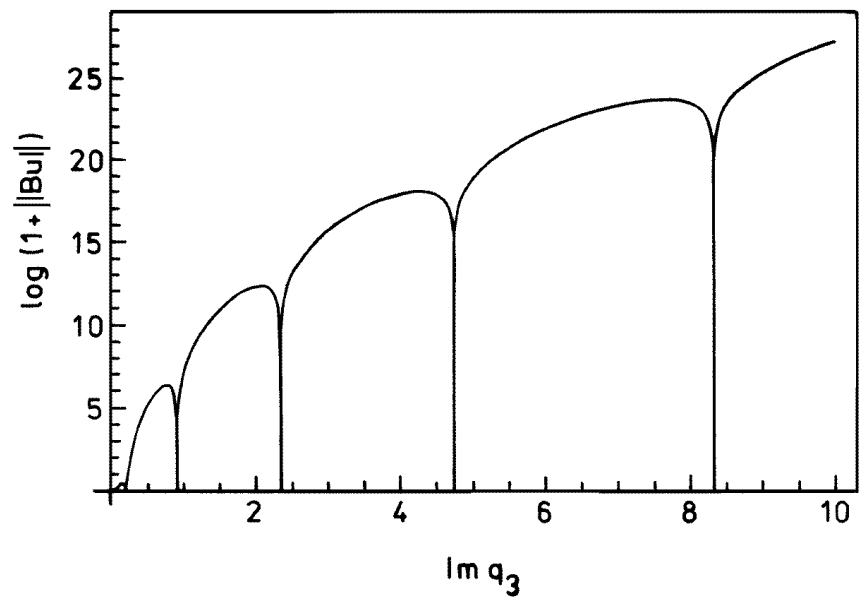}
\caption{Quantization of $\hat{q}_3$. Only half of the zeros shown constitutes the 
physical spectrum.  
}
\label{fig:f1}
\end{figure}

\begin{figure}[htb]
\epsfig{width=12cm, file=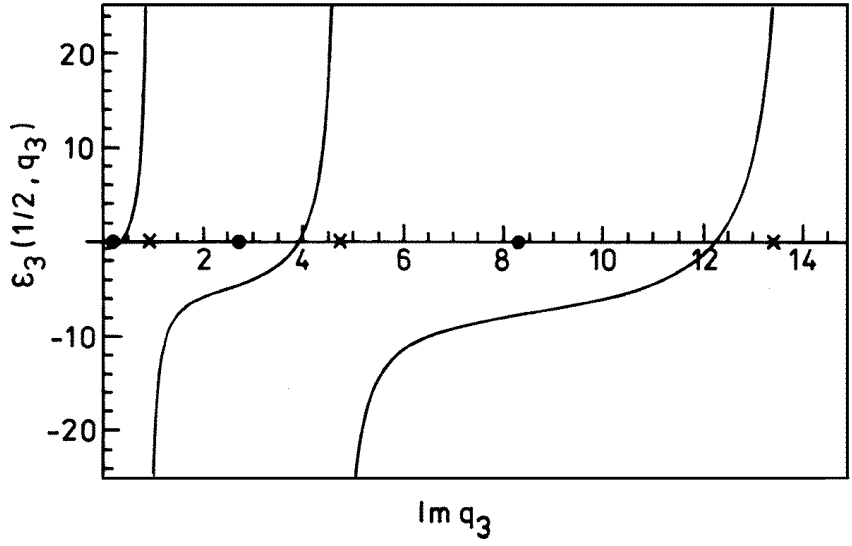}
\caption{ The holomorphic energy of the three reggeized gluons for imaginary
$q_3$ (solid lines \protect\cite{USPRL}). Dots and crosses show solutions of the first
 condition $(\ref{qond})$.
Solutions close to the poles of $\epsilon_3$ are eliminated by the second
 condition. 
}
\label{fig:f2}
\end{figure}

\end{document}